\documentclass[12pt,preprint]{aastex}

\shorttitle{NICMOS Observations of Shocked H$_2$ in Orion}
\shortauthors{Colgan et al.}

\begin{document}

\title{NICMOS Observations of Shocked H$_2$ in Orion$^1$}
\author{Sean W.J. Colgan, A.S.B. Schultz\altaffilmark{2}, 
M.J. Kaufman\altaffilmark{3}, E.F. Erickson, D.J. Hollenbach}
\affil{NASA-Ames Research Center, Moffett Field, CA 94035}

\email{sean.colgan@nasa.gov}
\altaffiltext{1}{
Based on observations made with the NASA/ESA {\it Hubble Space Telescope} 
obtained at the Space Telescope Science Institute, which is operated by the Association of 
Universities for Research in Astronomy, Inc., under NASA contract NAS 5-26555. }
\altaffiltext{2}{University New South Wales and SETI Institute}
\altaffiltext{3}{San Jose State University}

\begin{abstract}

HST NICMOS narrowband images of the shocked molecular hydrogen emission in OMC-1 
are analyzed to reveal new information on the BN/KL outflow. 
The outstanding morphological feature of this region is 
the array of molecular hydrogen ``fingers'' emanating from 
the general vicinity of IRc2 and the presence 
of several Herbig-Haro objects. 
The NICMOS images appear to resolve individual shock fronts. 
This work is a more quantitative and detailed
analysis of our data from a previous paper 
\citetext{Schultz et~al$.$}. 

Line strengths for the H$_2$ 1--0 S(4) plus 2--1 S(6) lines at 1.89 \micron\ are estimated 
from measurements with the Paschen $\alpha$ continuum filter F190N 
at 1.90 \micron, and continuum measurements at 1.66 and 2.15 \micron. 
We compare the observed H$_2$ line strengths and ratios 
of the 1.89 \micron\ and 2.12 \micron\ 1--0 S(1) lines with models for 
molecular cloud shock waves. 
Most of the data cannot be fit by J-shocks, 
but are well matched by C-shocks 
with shock velocities in the range of 20--45\,km\,s$^{-1}$ 
and preshock densities of $10^{4} - 10^{6}$ cm$^{-3}$, 
similiar to values obtained in larger beam studies which averaged over many shocks. 
There is also some evidence that shocks with higher densities have lower velocities. 

\end{abstract}


\section{Introduction}

At 450 pc, the Orion molecular cloud is the nearest and best-studied
region of massive star formation.  
The Trapezium stars, formed within Orion Molecular Cloud 1, 
have cleared a cavity at the near edge of the cloud. 
The visible Orion Nebula is the thin layer of photo-ionized gas 
on the cavity's surface facing the observer.  
Behind M42, and further from the observer, 
is a photodissociation region (PDR) also excited by the Trapezium.
The BN/KL region lies still deeper in the cloud, 
beyond both the ionized gas and the PDR. 
This region contains embedded sources  
with one or more associated outflows; 
the total luminosity of this region approaches $\sim 10^{5}L_{\sun}$ \citep{GS89}. 
The mid-infrared source IRc2 had long been thought to be
the origin of these outflows; 
but \citet{D93} resolved IRc2 into four sources, 
raising the possibility that none of them is sufficiently powerful 
to drive the observed outflows. 
\citet{MR95} suggested 
that the origin may be closer to the infrared source ``n'' \citep{LBLS82}, 
located $\approx$ 5\arcsec\ SW of IRc2. 
\citet{GGDNMT} detected extended emission from source n 
at wavelengths out to 22 \micron, 
but estimate a luminosity of only 2000 $L_{\sun}$. 
They also resolved IRc2 into $\sim5$ knots 
and suggested that these sources together with radio source I 
\citep{Ch87} comprise at least part of the core of 
a high density star forming cluster. 

The most striking of the molecular hydrogen outflows is an $\sim3'$ (0.4 pc) sized, 
butterfly-shaped region of H$_2$ emission, centered to the north of BN, 
which exhibits line ratios typical of shock excitation \citep{B78}.
From \ion{O}{1} emission, \citet{AT84} identified 
a number of optical HH objects in this vicinity. 
\citet{Ta84} discovered peculiar linear H$_2$
structures in the outflow. 
\citet[hereafter AB]{AB93} showed that these H$_2$ ``fingers'' 
and all the associated optical HH objects
at the far northern end of the outflow 
(approximately 120\arcsec\ from BN) 
terminated in knots of \ion{Fe}{2} emission. 
\citet{S98} found additional H$_2$ fingers within 30\arcsec\ of BN. 
Schultz et~al$.$ \citetext{1999, Paper 1}, 
found that only 2 of the 15 inner fingers seen by \citet{S98} 
had bow shocks capped by \ion{Fe}{2} emission, 
suggesting a lower excitation than in the outer fingers seen by AB. 

From offsets between the peak H$_2$ emission and the peak H$_2$ velocity, 
\citet*{GKCFLPRL} suggested that the H$_2$ emission arises in part 
from outflows from protostars within dense clumps of gas. 
In contrast, based on proper motion studies of optical features, 
\citet*{D02} found that both finger systems could have been created by an explosive event 
close to the IRc2-BN complex which took place approximately 1000 years ago. 
Interestingly, \citet*{R05} and \citet*{G05} suggested that BN and sources I and n were 
originally part of a multiple massive stellar system that disintegrated about 500 years ago. 
Explanations for the unique system of fingers have focused on two theories.  
AB originally suggested that they are ``bullets''---ejected clumps leaving
a wake of shocked material behind them.  
However, the observed morphology of the H$_2$ emission is inconsistent 
with models for bullets (\citealp{SN92}, \citealp*{KMC94}, \citealp{XS95}, \citealp*{JYT96}). 
These models predict that rapidly moving clumps are fragmented 
and also predict that the tails should be pointing away from the ejection source, 
which is not seen.
\citet*{SXM95} suggested the features
are produced when a faster wind collides with a slower, older outflow. 
Rayleigh-Taylor instabilities from the collision form the clumps in situ, 
moving at the speed of the older outflow. 
The observed fingers then condense behind the slowly moving clumps.
This is similar to the mechanism thought to have produced the cometary knots
in the Helix Nebula \citep{OH96}.  
One prediction of this model is that 
a region of clumpy H$_2$ emission will form behind 
(i.e. upstream of) the bullets. 
\citet{MML97} claimed to have found
this clumpy emission in the central region of the H$_2$ outflow.  
However, our previous work \citep{Paper1} 
and that of \citep{S98} 
shows that much of this \lq\lq clumpy" H$_2$ emission 
is resolved into more discrete objects, 
some resembling additional fingers.   
The remaining, unresolved clumpy emission 
is often mixed with the inner fingers. 

Here we discuss the interpretation 
of our previously published NICMOS infrared images of a 
90\arcsec\ wide region centered on BN/KL, 
focusing on the structure of the H$_2$ emission. 
Examination of the F190N images, 
originally obtained for subtracting continuum from 
the P$\alpha$ 1.87 \micron\ images \citep{Paper1}, 
suggested that in many regions there was a strong correlation
with the H$_2$ 1--0~S(1) 2.12 \micron\ continuum-subtracted images. 
In fact, the F190N filter bandpass includes both 
the H$_2$ 1-0~S(4) and the 2-1~S(6) lines at 1.89 \micron.
This H$_2$ emission is likely produced in shocks 
\citep{GFTL76}. 
H$_2$ emission can also be produced through UV fluorescence, 
but larger beam studies of multiple H$_2$ transitions by 
\citet{U96}, \citet*{R00}, and many others 
have shown that the measured line ratios 
in this region are consistent only 
with thermal excitation and not UV fluorescence. 
In this paper, we focus on the finger-like structures and the HH
objects.  Some of these objects have optical counterparts, which places 
constraints on their position within the cloud/nebula interface.  
In \S2 we discuss the observations and data reduction. 
In \S3 we compare the 
observed H$_2$ line brightnesses with several classes of shock models 
in order to determine shock types, shock velocities, and gas densities.  
In \S4 we discuss the morphology of,
and emission from, many of the more distinct, brighter features seen in our images. 

\section{Observations}

Observations were made of H$_2$ 1--0~S(1) 2.12 \micron\ and 
1--0~S(4) plus 2--1~S(6) 1.89 \micron\ in the 1\% bandpass NICMOS filters. 
The initial reduction of the NICMOS data was described in detail in Paper 1 
and generally followed standard procedures. 
Photometry was then performed on the reduced data from Paper 1 
using the IRAF \citep{T93} task {\bf polyphot}, 
in which the average brightness is estimated inside a user-defined polygonal aperture.  
The apertures were designed to closely follow the outline of each object we identified.  
Sky subtraction was not performed with the {\bf polyphot} task,
but separately using a region or regions far from areas of obvious emission.  
Regions selected for photometry are shown in Figure 1. 
Knot designations, except for HH 208, 
utilize the source identification scheme of \citet{OW94}. 
Knot identifications for HH 208 are shown in Figure 11.
The features were selected for being distinct and fairly bright, 
having detectable 1.66 \micron\ and 2.15 \micron\ 
continuum emission, and 2.12 \micron\ line emission. 
Knot U was also included even though it had no detectable 1.66 \micron\ continuum. 
The resulting photometry is listed in Columns 2-5 of Table 1. 
The formal statistical errors are such that the signal-to-noise ratios of all the 
measurements in Table 1 exceed 25 except for three: 
i) the 1.66 \micron\ measurement of 128--248 (S/N $= 4$), 
ii) the 1.66 \micron\ measurement of HH 208U (not detected), 
and 
iii) the 2.15 \micron\ measurement of HH 208U (S/N=15). 

There are no continuum observations for the H$_2$ 1.892 \micron\ 1--0~S(4) plus 2--1~S(6) images. 
However, we did observe continua at 1.66 and 2.15 \micron, 
intended as continua for the \ion{Fe}{2} and 1--0~S(1) lines respectively. 
The 1.89 \micron\ continuum has been estimated from these 
continuum measurements by linearly interpolating between 
the 1.66 \micron\ and 2.15 \micron\ photometric measurements. 
The continuum interpolation approach was checked by applying the same technique 
to nine regions with no 2.12 \micron\ H$_2$ emission. 
In these test regions, the interpolated continuum value was on average 
$99\pm3$\% of the value actually measured in the F190N filter, 
confirming that this is a reasonable approach.
As an example of the interpolation, the brightnesses for 128--248, 
with the estimated continuum is shown in Figure 2.

\clearpage
\begin{figure}
\figurenum{1}
\plotone{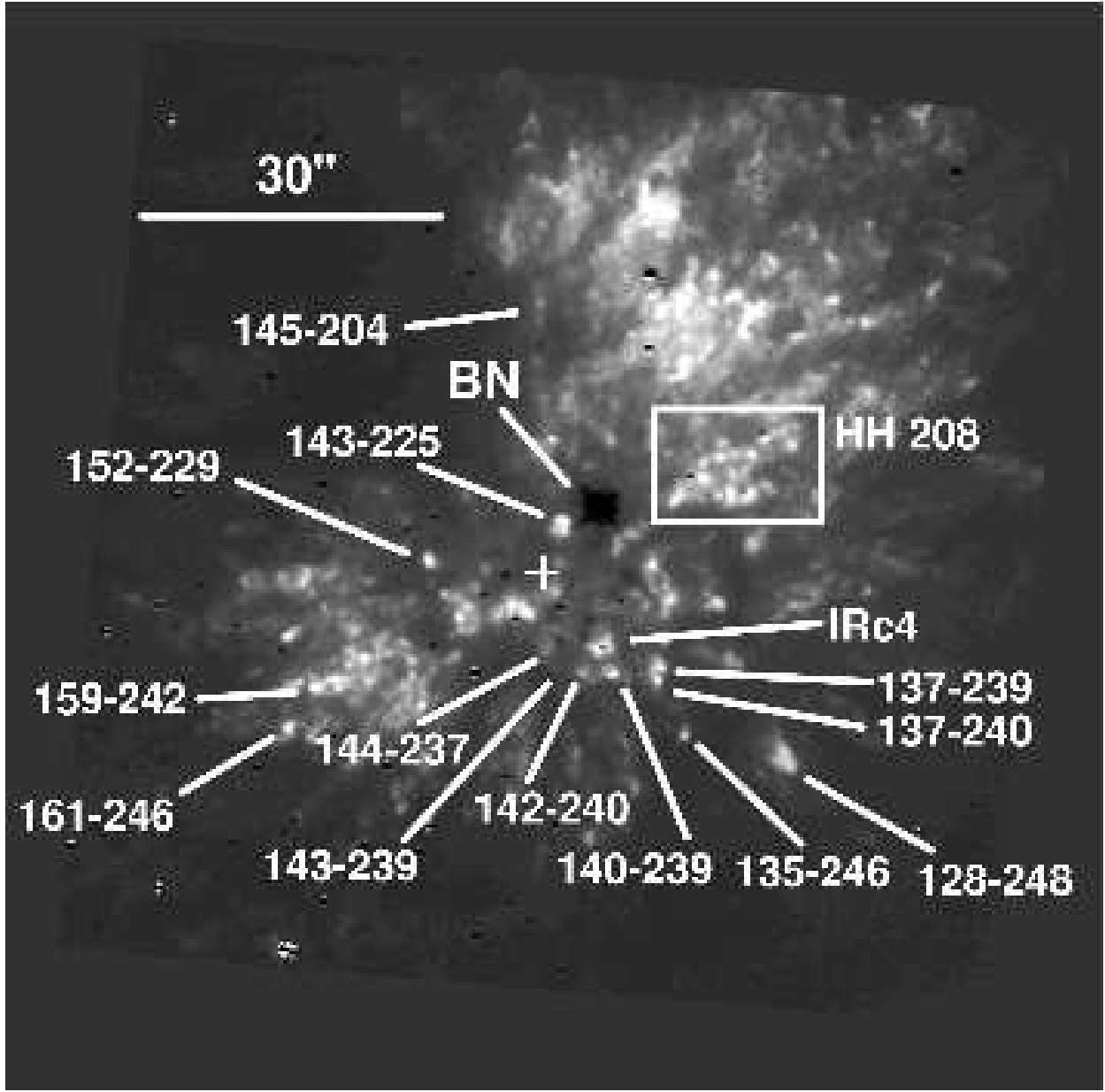}
\caption{H$_2$ 1--0 S(1) image of the Orion region with the selected features labelled. 
North is up and east is to the left. 
BN is located at $05^{\rm h}35^{\rm m}14.^{\rm s}12, -05^{\degr}22'23.''2$ (J2000) 
and the location of IRc2 is noted by a cross.
\label{fig1}}
\end{figure}
\clearpage
\begin{deluxetable}{lcccccccccc}
\tabletypesize{\small}
\tablecolumns{11}
\tablewidth{0pt}
\tablecaption{H$_2$ Photometry and Results\label{table1}} 
\tablehead{
\colhead{} && \multicolumn{4}{c} {Measured Line plus Continuum} 
&& \multicolumn{4}{c}{Extinction Corrected H$_2$ Brightnesses} \\
\colhead{Position} &\ \ & \colhead{1.66 \micron}
& \colhead{1.90 \micron} & \colhead{2.12\micron}
& \colhead{2.15 \micron}   &\ \ 
& \colhead{$A_{2.12 \mu{\rm m}}$\tablenotemark{a}}
& \colhead{$A_{V}$\tablenotemark{b}} 
& \colhead{1.89 \micron} & \colhead{2.12 \micron} \\
\colhead{} && \multicolumn{4}{c}
{[$10^{-3}$ ergs  s$^{-1}$ cm$^{-2}$ sr$^{-1}$]}
&& \colhead{(mag)} & \colhead{(mag)} & \multicolumn{2}{c}
{[$10^{-3}$ ergs  s$^{-1}$ cm$^{-2}$ sr$^{-1}$]} \\ 
}
\startdata
HH 208A  && 0.31 & 1.14 & 2.63 & 1.37 && 0.6 & 4.5 & 0.70 & 2.33  \\
HH 208B  && 1.15 & 3.90 & 11.1 & 3.87 && 0.6 & 4.5 & 2.94 & 13.0  \\
HH 208D  && 0.85 & 2.55 & 9.96 & 1.96 && 0.6 & 4.5 & 2.31 & 14.0  \\
HH 208E  && 0.72 & 2.28 & 7.96 & 1.84 && 0.6 & 4.5 & 2.03 & 10.8  \\
HH 208F  && 0.25 & 1.88 & 7.96 & 1.72 && 0.6 & 4.5 & 1.90 & 11.0  \\
HH 208J  && 0.54 & 2.28 & 9.49 & 1.84 && 0.6 & 4.5 & 2.24 & 13.5  \\
HH 208N  && 3.31 & 4.10 & 9.53 & 3.09 && 0.6 & 4.5 & 1.50 & 11.1  \\
HH 208P  && 0.02 & 1.68 & 7.66 & 1.01 && 0.6 & 4.5 & 2.38 & 11.7  \\
HH 208R  && 1.85 & 2.55 & 9.15 & 1.60 && 0.6 & 4.5 & 1.45 & 13.0  \\
HH 208U  && 0.0  & 0.42 & 1.82 & 0.21 && 0.6 & 4.5 & 0.63 & 2.82  \\
128--248 && 0.02 & 1.28 & 6.34 & 1.01 && 1.0 & 8   & 2.57 & 13.6  \\
135--246 && 0.33 & 1.08 & 3.20 & 0.71 && 0.6 & 4.5 & 1.10 & 4.38  \\
137--239 && 0.34 & 1.48 & 7.02 & 0.95 && 1.0 & 8   & 2.61 & 15.4  \\
137--240 && 0.33 & 1.54 & 7.32 & 1.01 && 1.0 & 8   & 2.76 & 16.0  \\
140--239 && 0.56 & 2.48 & 9.74 & 2.37 && 0.6 & 4.5 & 2.16 & 13.0  \\
142--240 && 0.54 & 1.81 & 5.28 & 1.72 && 0.6 & 4.5 & 1.43 & 6.34  \\
143--225 && 0.85 & 3.90 & 12.6 & 5.15 && 1.0 & 8   & 3.80 & 32.4  \\
143--239 && 0.51 & 0.94 & 1.49 & 0.83 && 0.6 & 4.5 & 0.54 & 1.19  \\
144--237 && 0.54 & 1.20 & 3.74 & 0.89 && 0.6 & 4.5 & 0.96 & 4.98  \\
145--204 && 0.18 & 1.01 & 4.17 & 1.13 && 1.0 & 8   & 1.31 & 7.83  \\
152--229 && 0.30 & 2.15 & 7.83 & 3.15 && 0.6 & 4.5 & 1.16 & 8.51  \\
159--242 && 0.69 & 3.22 & 12.0 & 1.84 && 0.6 & 4.5 & 3.88 & 17.8  \\
161--246 && 0.85 & 2.48 & 7.70 & 1.54 && 0.6 & 4.5 & 2.52 & 10.8  \\
\enddata
\tablenotetext{a}{Extinction estimates from \citet{Ch97} and \citet{R00} - see text for more discussion.}
\tablenotetext{b}{Conversion to $A_{V}$ based on \citet*{C89}}
\end{deluxetable}

\clearpage
\begin{figure}
\figurenum{2}
\plotone{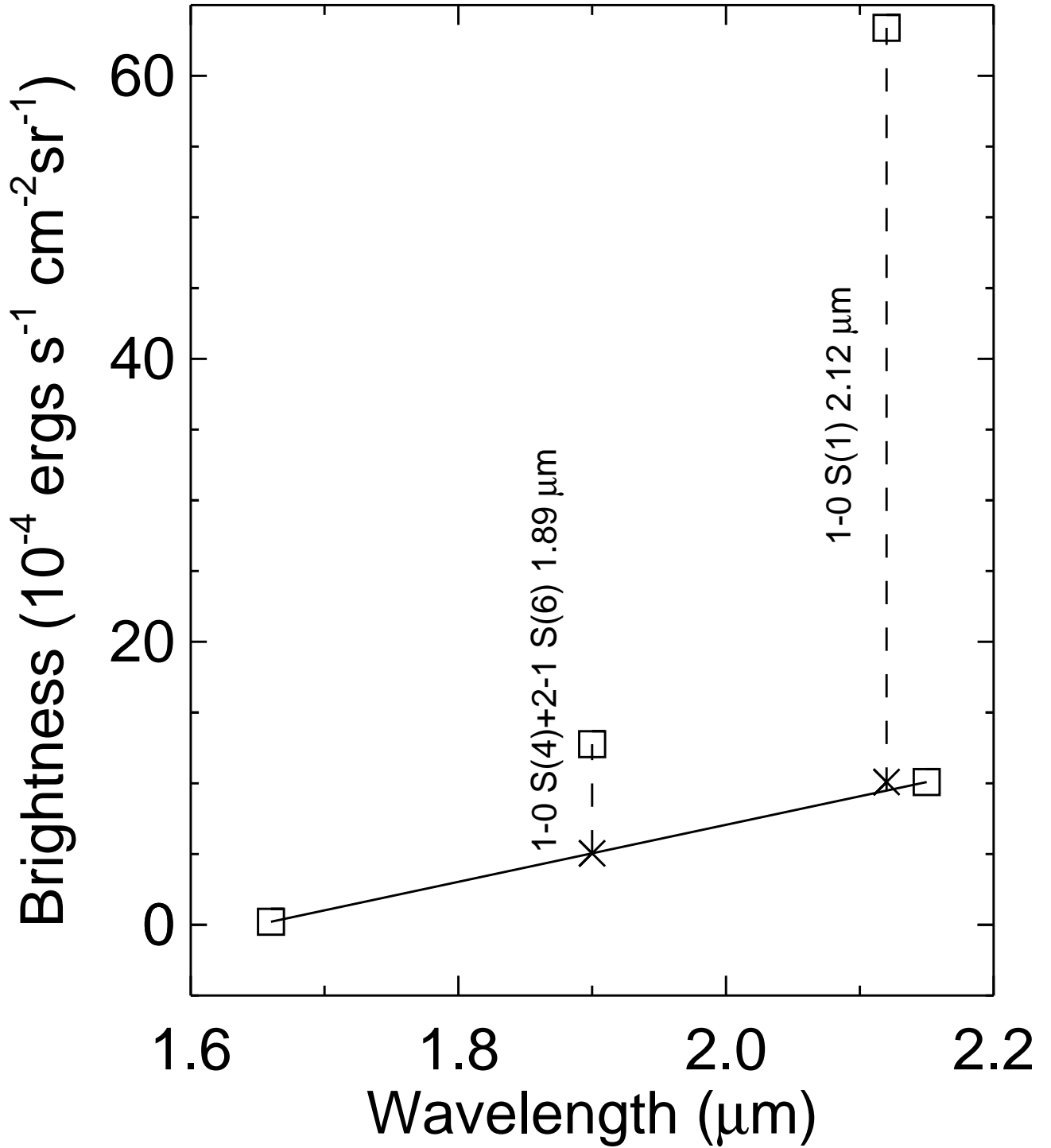}
\caption{The estimation of H$_2$ line brightnesses for 128--248. 
The boxes show the measured photometric points; errors are smaller than the boxes. 
The X's show the estimated continuum at the wavelengths of the H$_2$ lines. 
The 2.12 \micron\ continuum was set equal to the 2.15 \micron\ measurement. 
\label{fig2}}
\end{figure}
\clearpage
The extinction corrections in this region are complex, and vary depending
upon the distribution of intervening dust, 
which lies not only within the molecular cloud but also 
in the PDR and in foreground material. 
Extinction estimates are also complicated by reflection off the
back side of the nebula. 
Most of our objects, 
including two fingers and several more compact structures, 
have some associated \ion{Fe}{2} (1.64 \micron) emission 
\citep{Setal07}. 
For these objects, the 2.12 \micron\ extinction can be estimated from 
the results of \citet{Ch97}, 
who used measurements of the \ion{Fe}{2} 1.257 and 1.644 \micron\ transitions, 
and the extinction curve of \citet*{C89}, 
to obtain the extinction values of 0.6 shown in column 6 of Table 1. 
For five regions, no \ion{Fe}{2} emission is seen, possibly because of larger extinction. 
The extinction for these particular objects hasn't been estimated 
by \citet{Ch97}, or others. 
The best references are large beam extinction studies 
of the H$_2$ Peak 1 source. 
The largest and most recent such study is that of 
\citet{R00}, 
based on the ISO measurement of 
56 H$_2$ transitions covering wavelengths of 2-17 \micron. 
They find A$_K=1.0\pm0.1$, a value we adopt here for  
128--248, 137--239, 137--240, 143--225, and 145--204. 
For wavelengths other than 2.12 \micron, the brightnesses have been extinction 
corrected using A$_{\lambda} =$ A$_{2.12}\{{{\lambda}\over{2.12}}\}^{-1.61}$ \citep*{C89}, 
who also found that the shape of the extinction curve in the IR 
is independent of the value assumed for R$_V$. 
Usage of different plausible extinction corrections, 
within the uncertainties, does not appreciably alter our conclusions.
For reference, we show in column 7 the approximate visual extinctions, A$_V=4.5$ 
and A$_V\sim8$, based upon R$_V=$A$_V$/E$_{(B-V)}=3.1$, and \citet{C89}. 
The final 1.89 \micron\ H$_2$ line-to-continuum ratios 
range from 0.25 to 1.1, with a median of 0.7. 

\section{Shocked Emission Features: Line Ratio Analysis}

Early shock models of the Orion outflow invoked planar C-type shock models
to explain the emission from species such as H$_2$ and CO 
(e.g.~\citealp{DR82}, \citealp*{CMH82}). 
C-type or ``continuous'' shocks occur at relatively low shock speeds ($V_{shock}
\lesssim 50$\,km\,s$^{-1}$) 
in the presence of magnetic fields. 
A low ionization fraction allows ionized gas 
to cushion the shock in the neutral gas, 
limiting the neutral gas temperature to 
less than several thousand Kelvin and preventing 
significant dissociation. 
J-type or ``jump'' shocks generally occur 
at relatively high shock speeds ($V_{shock} \gtrsim 50$\,km\,s$^{-1}$) 
and usually dissociate molecular gas in 
the high temperature ($T \sim 10^4-10^5$ K) post-shock region. 
Molecules reform in the cooling post-shock gas at T$\sim$500~K.
For pre-shock densities $\gtrsim10^5\,\rm cm^{-3}$,
H$_2$ line ratios produced in the 
reforming molecular gas may reach values higher than thermal values since
H$_2$ reforms in excited states, leading to a non-thermal 
cascade through rovibrational states \citep{HM89}.

Observations of shocked 
H$_2$O emission in Orion \citep[e.g.][]{HNMK98} 
seem to confirm the general picture 
that C-type shocks are responsible for the molecular emission in the outflow.
These studies converge on preshock conditions 
$n(\rm H_2)\sim 10^5\,\rm cm^{-3}$ and 
$V_{shock}\sim 35$\,km\,s$^{-1}$ \citep[e.g.][]{CMH82}. 
It should be noted,
however, that these studies fit data collected from a beam area 
covering an entire outflow lobe ($\sim 1\arcmin \sim 0.1$ pc at Orion). 
Images of the shocked emission on sub-arcsecond 
scales \citep[AB,][]{S98} 
show many emission features, 
each of which presumably has its own shock conditions. 
What is unique about the NICMOS observations of the inner region 
is that the 0.\arcsec 2 ($\sim 1.4\times10^{15}\,\rm cm$) resolution 
allows the isolation of individual shock fronts on the length scales
expected for such shocks. 
These scales are expected to be $\sim 10^{15}\,\rm cm$, 
depending on the preshock density but with only 
a weak dependence on the shock velocity \citep{KN96}. 
This means that our deduced shock parameters are more 
likely to represent the local physical conditions, 
rather than an average over a number of shocks.

H$_2$ adaptive optics observations of the ambient molecular cloud 
to the south-east of BN/KL 
have been carried out with $0.''15$ angular resolution \citep{V01,kgfclvp}. 
The observed H$_2$ 1-0 S(1) brightness is 
well matched by C-shock models with shock velocities of 30\,km\,s$^{-1}$ 
and pre-shock densities of $10^6\,\rm cm^{-3}$, 
but the same models fall short 
of matching the 2-1 S(1) brightness by a factor $\sim$2 \citep{V01}. 
Higher shock velocity models improve the 2-1 S(1) brightness prediction, 
but provide a worse fit to the 1-0 S(1) brightness. 
J-shock models produce the observed 2-1 S(1)/1-0 S(1) brightness ratio but have 
trouble reproducing the individual line brightnesses. 
\citet{P01} suggested that non-stationary C-shocks can reproduce the high brightness 
and the large observed 2-1 S(1)/1-0 S(1) brightness ratios. 
However, non-stationary C-shocks have difficulty accounting for 
the proper motion velocity measurements of \citet{D02}, 
who found that over the velocity range of 20 -- 400\,km\,s$^{-1}$, 
different emission lines from the same object have similar velocities. 
From comparison with the higher excitation H$_2$ 2-1 S(1) line, 
\citet*{kgfclvp} found H$_2$ clumps to have abrupt, south-facing edges 
exhibiting high excitation temperatures. 
Even C-shocks propagating into high density material can't account for 
the excitation temperature maxima and the line fluxes. 
This led \citet*{kgfclvp} to suggest that J-shocks propagating into 
material with a pre-shock density $\ge 10^6\,\rm cm^{-3}$, 
plus an additional contribution from photo-dissociated material, 
are required to explain these measurements. 
Emission from bow-shaped shock fronts \citep{S91}, 
in which both C-shocks and 
J-shocks are responsible for the overall emission, 
may also help explain these observed brightnesses and line ratios. 

For our study, 
we use the ratio of the H$_2$ line fluxes at 2.12 and 1.89 \micron\ 
as a diagnostic of conditions in the emitting gas. 
The Camera 3 F190N filter transmission plot was examined to determine 
the relative throughput of the 1-0 S(4) (1.892 \micron ) and 
2-1 S(6) (1.8947 \micron ) rovibrational transitions.
From the transmission plot, we estimate the NICMOS filter transmissions 
at the H$_2$ wavelengths to be 64\% and 98\% respectively. 
Hence, we compare the measurements with 
model predictions for $0.64\times$B[1-0 S(4)]$+ 0.98\times$B[2-1 S(6)]. 
All three transitions require gas temperatures in excess 
of 1000 K in order to produce significant emission. 
Such high temperatures imply that shock excitation
is responsible for the H$_2$ emission. 
A difficulty with interpreting the 2.12 and 1.89 \micron\ lines is that 
the transitions are relatively close together in excitation:
the upper state of the S(1) line is at 6956K, 
and that of the S(4) and S(6) lines are at 9286K and 16880K, 
which can be a rather small baseline to fit. 

In an effort to characterize the shocked H$_2$ emission, 
we have compared our observed line brightnesses and line ratios 
with standard models of shock emission: 
the C-shock model of \citet{KN96} 
and the J-shock model of \citet{HM89}. 
We first consider 128--248, a well-isolated source 
for which we obtain the highest signal-to-noise ratio 
at 1.89 and 2.12 \micron. 
Because the large beam studies suggest shocked gas with densities near $10^5\,\rm cm^{-3}$, 
we computed the brightnesses of the 1-0~S(1) and $0.64\times$B[1-0 S(4)]$+ 0.98\times$B[2-1 S(6)] 
transitions in C-shocks with preshock densities of $10^4,\,10^5$ and $10^6\,\rm cm^{-3}$ and 
shock speeds up to 50\,km\,s$^{-1}$; 
and in J-shocks with preshock densities of $10^5,\,10^6$ and $10^7\,\rm cm^{-3}$ and 
shock speeds up to 100\,km\,s$^{-1}$.
In line with current practice, the models assume 
a) a planar shock, 
b) a magnetic field perpendicular to the direction of propagation of the shock, and 
c) the magnetic field strength in microgauss equals the square root 
of the density in cm$^{-3}$ \citep{TH86}. 
The precise structure of the C-shocks depends on the field strength, 
as well as the ionization fraction, the grain size distribution, and  
the details of gas cooling. 
We have previously explored the effects of varying these parameters  
on shock structure \citep{KN96}. 
We find that while the precise value of shock velocity and density determined
from a line ratio may vary from those presented here, 
the range of intensities and line ratios possible in C-shock models 
is essentially limited by the temperature at which H$_2$ dissociates. 
Thus our conclusion that C-shocks can explain the emission in most of the observed features 
is robust even if the precise shock parameters are different from those we have assumed.

At this assumed magnetic field strength, low velocity shocks are C-shocks and 
not the lower magnetic field strength, non-dissociative, molecular J-shocks discussed by \citet{W00}. 
Typically, molecular J-shocks produce a factor of 10-1000 times fainter H$_2$ emission 
in the lines discussed here than C-shocks \citep{W00}. 
Since these line fluxes would be undetectable, 
we have not included non-dissociative, molecular J-shocks in our grid of models. 
Higher velocity shocks would be dissociative J-shocks, which we do consider.
C-shocks are also known to be unstable to the Wardle instability \citep{W90} 
arising from perturbations in the magnetic field direction,
an effect which is not taken into account in the steady-state C-shock models presented here. 
The effects of this instability on the strengths of H$_2$ emission lines 
has been explored by \citet{MS97} and \citet{NS97}.   
Both studies reached the conclusion that, for shocks over the range
of densities and shock velocities we consider 
(i.e. shocks for which H$_2$ is the dominant coolant), 
the instability has little effect on the predicted intensities of H$_2$ lines. 
Thus our steady-state models should be sufficient for modeling the emission presented here.
Clearly, changing the assumptions in the shock models - 
bowshocks instead of plane-parallel shocks, 
stronger or weaker magnetic fields, 
a different orientation of the magnetic field with respect to the shock, etc. - 
will result in different predictions for the H$_2$ line strengths, 
perhaps resulting in better (or worse) agreement with our data. 
A complete exploration of this parameter space is beyond the scope of this paper, 
but we can show that for some commonly-assumed magnetic field properties, 
plane-parallel C-shock models provide a reasonable fit to the data, 
while plane-parallel J-shock models generally do not. 

The predicted brightness ratio \{$0.64\times$B[1-0 S(4)]$+ 0.98\times$B[2-1 S(6)]\}/B[1-0 S(1)] 
from each model calculation is shown in Figure 3 as a function of shock velocity. 
Also shown is the measured ratio, 0.19, for 128--248. 
The line ratio is consistent with either C- or J-shocks. 
However, the line ratio and brightness from this feature is best fit
by a C-shock model with velocity 36\,km\,s$^{-1}$ 
and preshock density of $6\times 10^4\,\rm cm^{-3}$. 
The brightnesses, line ratio, and model predictions 
for 128--248 are listed in Table 2. 
The model values are well within
the systematic measurement uncertainties. 
{\it J-shocks produce absolute intensities which are too low to match the observed values.}

\clearpage
\begin{deluxetable}{cccc}
\tablecolumns{4}
\tablewidth{0pt}
\tablecaption{Observed Brightnesses and Model Fit for 128--248\label{table2}}
\tablehead{
& \colhead{$0.64\times$B[1-0 S(4)]$+ 0.98\times$B[2-1 S(6)]} & \colhead{B[1-0 S(1)]} & 
\colhead{$0.64\times$B[1-0 S(4)]$+ 0.98\times$B[2-1 S(6)]} \\
& \multicolumn{2}{c} {[$\rm erg\,s^{-1}\,cm^{-2}\,sr^{-1}$]} & /B[1-0 S(1)] \\
}
\startdata
Observed               & $2.57\times 10^{-3}$ & $1.36\times 10^{-2}$ & 0.190 \\
Model\tablenotemark{a} & $2.60\times 10^{-3}$ & $1.31\times 10^{-2}$ & 0.198 \\
\enddata
\tablenotetext{a}{Model parameters: $v_S=36$\,km\,s$^{-1}$ and 
$n=6.3\times 10^4\,\rm cm^{-3}$, which corresponds to an assumed magnetic field of 0.25 milli-Gauss.}
\end{deluxetable}

\clearpage
\begin{figure}
\figurenum{3}
\plotone{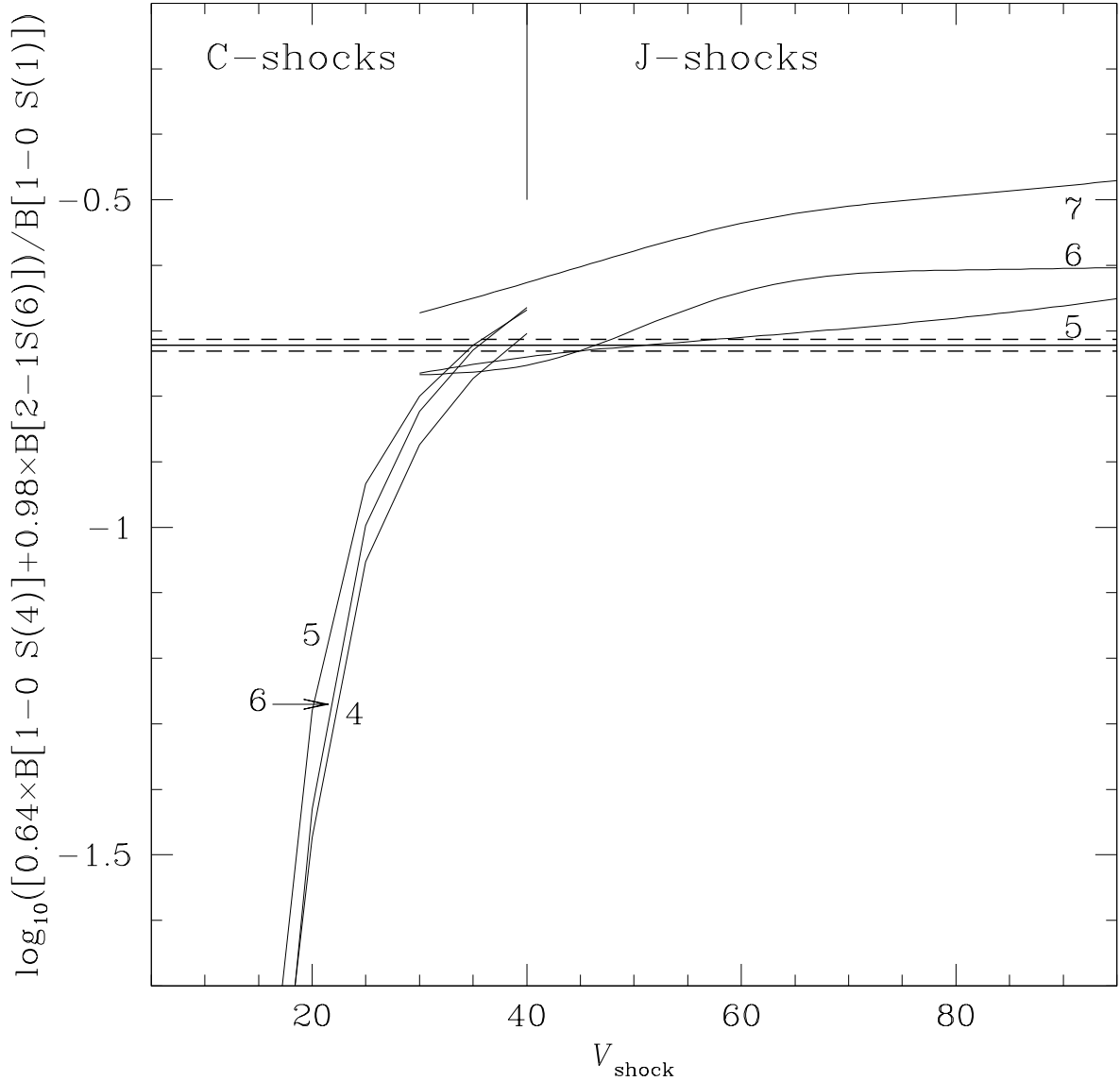}
\caption{Predicted \{$0.64\times$B[1-0 S(4)]$+ 0.98\times$B[2-1 S(6)]\}/1-0 S(1) line ratio 
from C-shock and J-shock models
with magnetic fields oriented perpendicular to the shock propagation direction 
and strengths in microgauss equal to the square root of the density 
in cm$^{-3}$ \citep{TH86}. 
Results are shown for C-shock models with $n(\rm H_2)=10^4,\,10^5$ and $10^6
\,\rm cm^{-3}$ and $V_{shock}=15-50$\,km\,s$^{-1}$, and  
J-shock models with $n(\rm H_2)=10^5,\,10^6$ and 
$10^7\,\rm cm^{-3}$ and $V_{shock}=30-100$\,km\,s$^{-1}$. 
Also shown is the measured value of the line ratio for 
the Orion feature 128--248 (horizontal solid line), 
with the statistical uncertainty in the 
ratio indicated by the dashed lines.
\label{fig3}}
\end{figure}

\clearpage
The extinction-corrected line flux ratio for each of 
ten locations in HH 208 and thirteen other features 
is plotted versus their 2.12 \micron\ H$_2$ line brightness in Figure 4. 
The shock model curves are overlain. 
Within the systematic measurement uncertainties, 
all but one of the observed line ratios 
are consistent with C-shocks having preshock densities $10^4\,-\,10^6\,\rm cm^{-3}$ and 
shock velocities of 20 to 45\,km\,s$^{-1}$. 
The narrow range of shock velocities is not surprising. 
Slower C-shocks produce much weaker H$_2$ emission and would not have been detected. 
Faster C-shocks break down into J-shocks, 
again with much fainter H$_2$ emission because the H$_2$ is dissociated. 
The consistency with larger beam studies does suggest that these studies 
yield reasonable average shock parameters. 
Table 3 lists shock velocities and pre-shock densities for all the objects. 
These values were estimated by interpolating within 
the grid of calculated C-shock models at 
densities of $10^4$, $10^5$, $10^6$ 
and shock velocities of 20, 25, 30, 35, and 40 \,km\,s$^{-1}$, 
whose curves are plotted in Figure 4. 
Extrapolations for those objects just outside the grid were made using 
a few additional models with shock velocities of 45 \,km\,s$^{-1}$. 
Two objects - 143--239 and HH208A - are clearly outside the C-shock grid. 

Examination of Figure 4 suggests that higher shock velocities 
may be correlated with lower pre-shock densities. 
A variety of statistical tests show that this correlation is significant at the 3-4$\sigma$ level. 
The log of the pre-shock density falls by $\sim 0.6$ 
for each shock velocity increase of 10 \,km\,s$^{-1}$. 
In order to search for structure 
such as the ``hot edges'' found by \citet*{kgfclvp},
we have constructed and examined images of the line flux ratio 
for all twenty three of the features included in Figure 4. 
We find no significant structure in the ratio images except for 145--204 and 159--242, 
where there is a peak in the line ratio offset by 1-2\arcsec\ 
from the peak H$_2$ 1--0 S(1) emission. 

\clearpage
\begin{figure}
\figurenum{4}
\includegraphics[clip, scale=0.75] {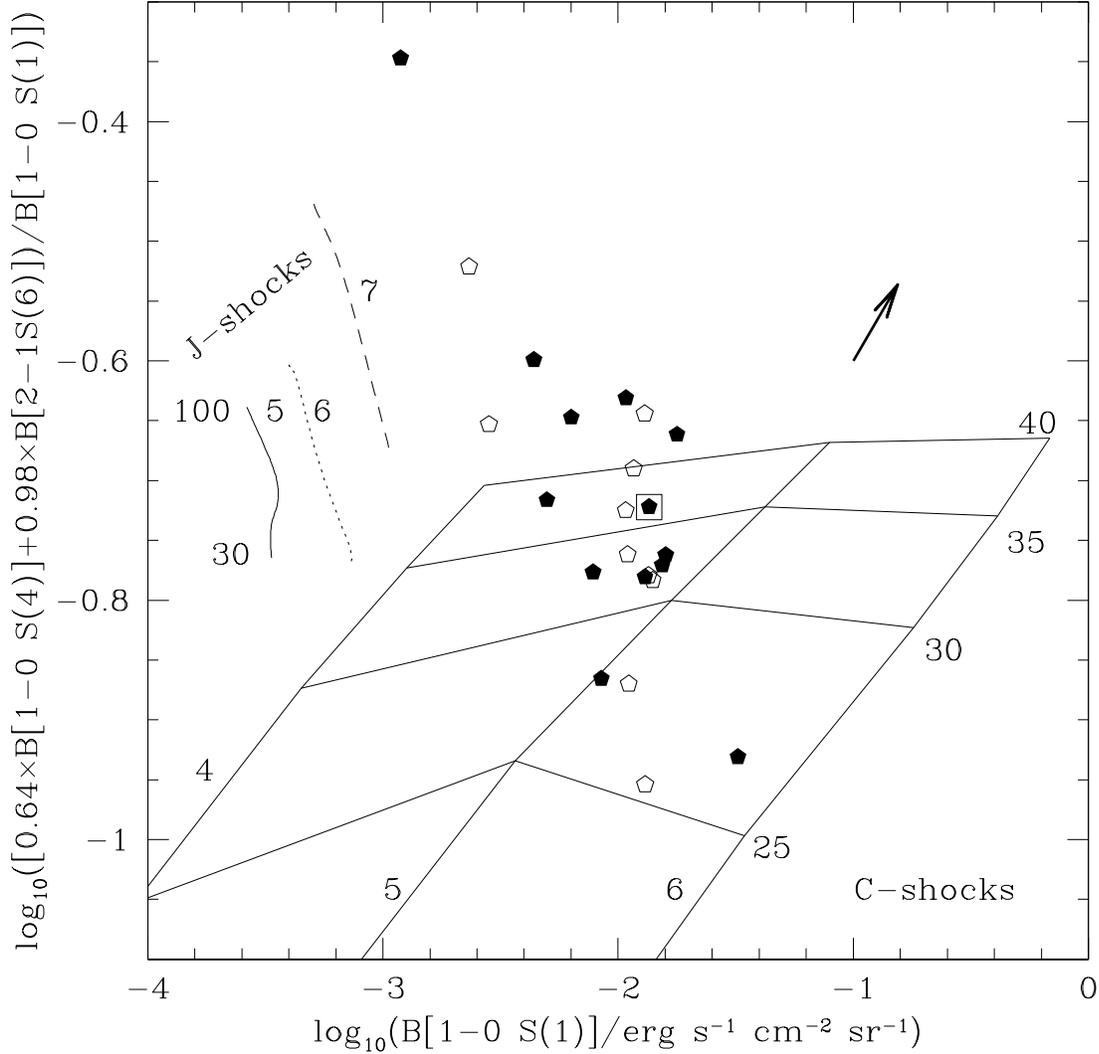}
\caption{The log of the \{$0.64\times$B[1--0 S(4)]$+ 0.98\times$B[2--1 S(6)]\}/[1--0 S(1)]
brightness ratio plotted versus the log of the 1--0 S(1) brightness. 
Measured values for each of the shocked fingers with both 1.89 \micron\ 
and 2.12 \micron\ emission are given by the pentagons. 
The symbols for the HH 208 positions are hollow.
The error bars in the data are smaller than the data symbols. 
A trend arrow indicating the direction and magnitude 
of the change in the observed ratio and brightness  
if the 2.12 \micron\ extinction was increased from 0.6 to 2.4 mag, 
is given by the solid arrow.
The symbol outlined by a box represents source 128--248. 
Model curves are shown for C-shocks 
(solid curves, labeled with log[$n(\rm H_2)$/cm$^{-3}]=$4, 5, 6 and 
shock velocity of 25--40\,km\,s$^{-1}$) 
and J-shocks (solid, dotted, and dashed curves, 
labeled with log[density/cm$^{-3}]=$5, 6, 7 and 
shock velocity of 30--100\,km\,s$^{-1}$). 
For both C- and J-shocks, 
the magnetic fields are oriented perpendicular to the shock propagation direction 
with strengths in microgauss equal to the square root of the density 
in cm$^{-3}$ \citep{TH86}. 
Factors contributing to uncertainties in the model predictions are discussed in \S3. 
\label{fig4}}
\end{figure}

\clearpage
\begin{deluxetable}{lcc}
\tabletypesize{\small}
\tablecolumns{3}
\tablewidth{0pt}
\tablecaption{Estimated C-Shock Velocities and Pre-Shock Densities \label{table3}} 
\tablehead{
\colhead{Position} & \colhead{$v_S$ (km\,s$^{-1}$)\tablenotemark{a}} & 
\colhead{log$[n_{\rm H_2}({\rm cm}^{-3})]$\tablenotemark{a}} \\
}
\startdata
HH 208B   & 42 & 4.5 \\
HH 208D   & 32 & 4.8 \\
HH 208E   & 36 & 4.5 \\
HH 208F   & 33 & 4.6 \\
HH 208J   & 32 & 4.8 \\
HH 208N   & 27 & 5.2 \\
HH 208P   & 40 & 4.5 \\
HH 208R   & 26 & 5.5 \\
HH 208U   & 42 & 4.0 \\
128--248  & 36 & 4.8 \\
135--246  & 45 & 4.0 \\
137--239  & 32 & 4.8 \\
137--240  & 33 & 4.8 \\
140--239  & 32 & 4.7 \\
142--240  & 42 & 4.3 \\
143--225  & 27 & 5.8 \\
144--237  & 38 & 4.3 \\
145--204  & 33 & 4.6 \\
152--229  & 27 & 5.0 \\
159--242  & 41 & 4.5 \\
161--246  & 42 & 4.4 \\
\enddata
\tablenotetext{a}{Assumes planar shock propagating into a perpendicular magnetic field with strength 
in microgauss equal to the square root of the density in cm$^{-3}$ \citep{TH86}.}
\end{deluxetable}

\clearpage
\section{Results and Discussion}

In this section, we outline the morphological and emission characteristics 
of the H$_2$ features. We categorize the features into fingers - 
structures which either exhibit a clear bow shock morphology 
or features we believe to be bow shocks approaching us at a low angle - 
and knots - mostly including a variety of clumps in HH 208. 

\subsection{Fingers}

The array of inner fingers which comprises the butterfly-shaped H$_2$ emission
first found by \citet{B78}, extends over a 90\arcsec\ broad region. 
Most of the northern H$_2$ fingers (AB) are outside of our field to the north, 
but there are additional fingers to the south, east of the
Trapezium \citep{MML97}. 
The velocity measurements of \citet{Ch97} 
found that in addition to strong, broad H$_2$ emission over the entire source, 
there are high velocity components confined to discrete condensations. 
The high velocity components are ascribed to additional `bullets' 
similar to those imaged in the northern fingers by AB. 
At our higher angular resolution, 
the morphology of the inner fingers (Finger 1) is quite varied.  
Some of the objects are revealed to be bright, well-defined 
bow shocks; others do not display distinct bows of any kind.  
Most of the objects are found to have complex structure, 
with what appear to be internal shocks.

\subsubsection{128--248 and 135--246}

128--248 is a bright, well-defined bow-shock in the southwestern portion of the finger array.  
Because it is so bright, and well-separated from the main array by dark dust lanes, 
128--248 is an excellent candidate for many studies. 
\citet{SB07} found a FWHM of 30 km\,s$^{-1}$, 
while \citet[their object 10]{Ch97} found it to have a FWZI of 150 km\,s$^{-1}$ . 
\citet{SB07} and \citet[their region 11]{GKCFLPRL} 
found that its line profile has no secondary line peaks, 
which is suprising for a bow shock, 
and almost unique among the inner fingers.
From its peak velocity of 0 km\,s$^{-1}$, \citet{GKCFLPRL} concluded that 128--248 
is moving in the plane of the sky, 
but were unable to determine in what direction.
The H$_2$ emission morphology is shown in Figure 5 - 
a classic bow-shock shape which appears to be moving to the southwest. 
From the line fluxes, we deduce a shock velocity and pre-shock density of 
36 km\,s$^{-1}$ and $10^{4.8}$ cm$^{-3}$ for 128--248. 

\clearpage
\begin{figure}
\figurenum{5}
\plotone{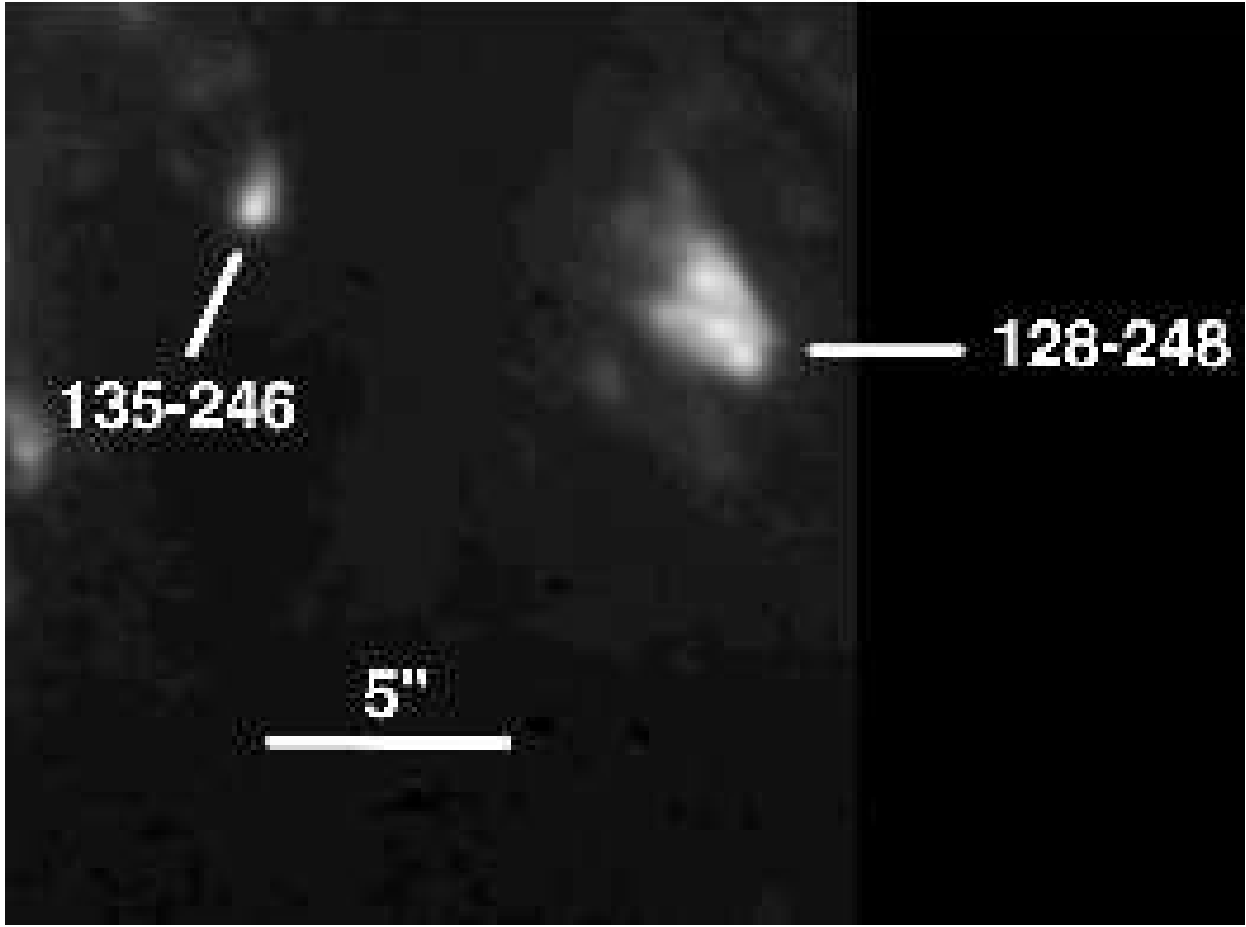}
\caption{H$_2$ 1--0 S(1) image of 128--248 and 135--246. North is up and east is to the left. 
The maximum brightness is 8.9$\times10^{-3}$ ergs s$^{-1}$cm$^{-2}$sr$^{-1}$. 
\label{fig5}}
\end{figure}

\clearpage
\begin{figure}
\figurenum{6}
\plotone{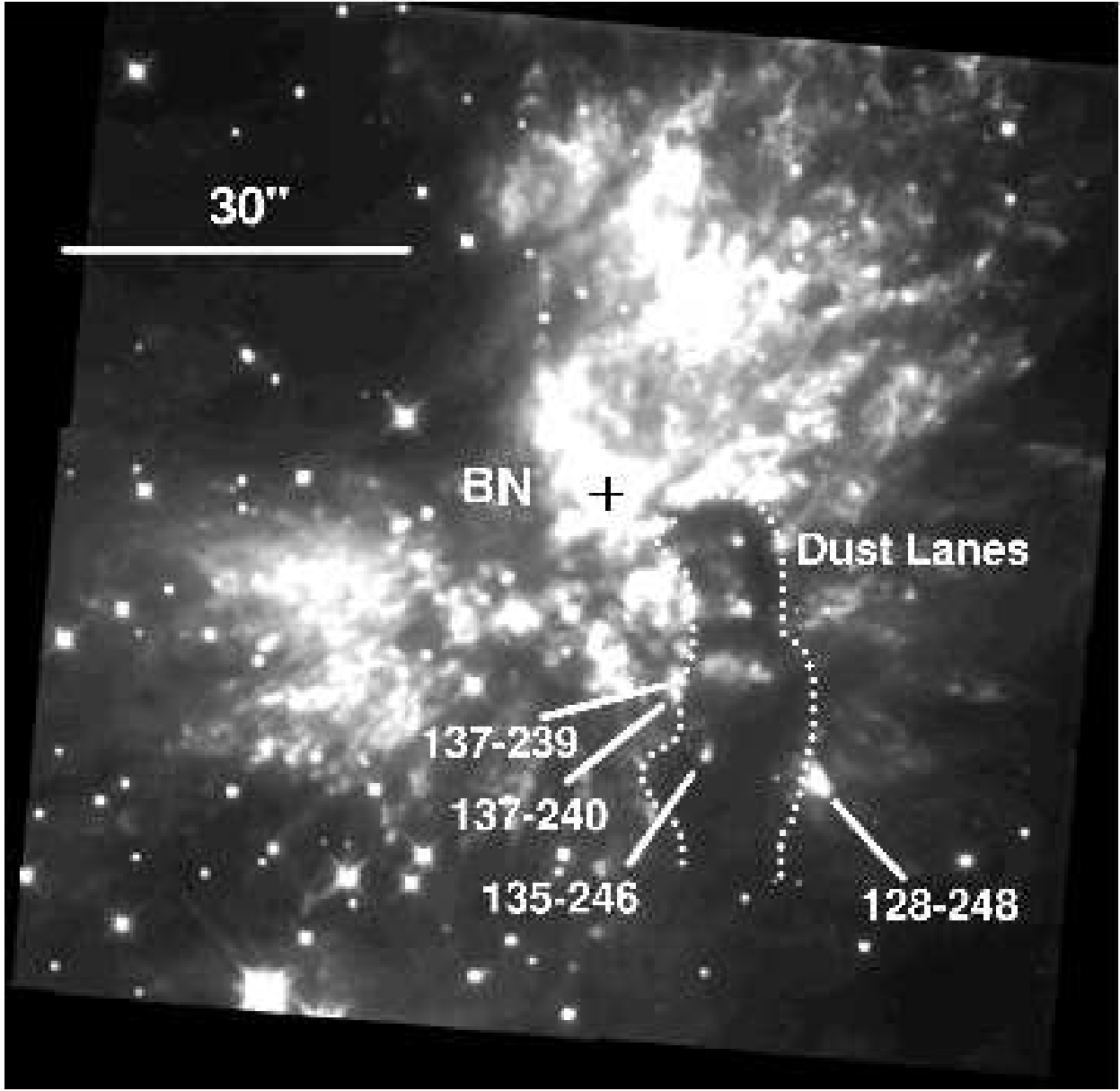}
\caption{2.12 \micron\ continuum plus H$_2$ 1--0 S(1) line map showing the locations 
of 128--248, 135--246, 137--239, and 137--240 
relative to the dust lanes and BN. The Trapezium is off the bottom of the image. 
North is up and east is to the left.
\label{fig6}}
\end{figure}
\clearpage

135--246 is a bright bow shock at the end of a faint finger 
emerging from a extended continuum ``tail'' of IRc4 (cf. Figure 1).  
It is prominent in H$_2$ 1-0~S(1), 
\ion{Fe}{2} \citep{Setal07}, and the F190N and F215N continuum bandpasses.  
The object also appears in \ion{O}{1} and possibly in \ion{S}{2} emission \citep{Setal07}. 
Our H$_2$ image is shown in Figure 5 - from the line ratios 
we deduce a shock velocity and pre-shock density of 
45 km\,s$^{-1}$ and $10^{4.0}$ cm$^{-3}$. 
The northeast boundary of the bright tip of the 135--246 bow shock is very sharp. 
We suggest this is probably due to the emergence of the finger from behind a 
region of high extinction:  
Figure 6 shows that the southwestern extent of
the H$_2$ finger is partially obscured by 
the same dark, curving dust lanes which are near 128-248 and the region south of BN. 

\subsubsection{137--239, 137--240, and 140--239}

The H$_2$ emission from 137--239, 137--240, and 140--239 is shown in Figure 7. 
From our analysis, 137--239, 137--240, and 140--239 are all fit by shock velocities of 
32 to 33 km\,s$^{-1}$ and pre-shock densities of $10^{4.7-4.8}$ cm$^{-3}$. 
140--239 is a small knot 3$''$ south of IRc4. 
NICMOS H$_2$ images \citep{S98} suggest that this object may be a bow shock, 
perhaps coming towards us at a low angle 
because of its relatively large blue-shift \citep{Setal07}.
The lower spatial resolution of \citet{Ch97} combined 137--239 and 137--240 
into a single feature, 
which they identified as a high-velocity ``bullet'' (\#5 in their list) with a FWZI of 100 km\,s$^{-1}$. 
The two features were also observed together in H$_2$ 1-0 S(0) as object b of \citet{L04}. 
The morphology in that line was essentially identical to the morphology in 1-0 S(1).  
\citet{GKCFLPRL} found these objects to have a peak velocity of -25 km\,s$^{-1}$, 
although they do not indicate whether they were able to distinguish between the two bow-shocks. 
\citet{SB07} find a velocity closer to -50 km\,s$^{-1}$ for 137--239. 
From Figure 6, we suggest 137--239 and 137--240 may also be emerging from a dust lane, 
like 135--246. 
\clearpage
\begin{figure}
\figurenum{7}
\plotone{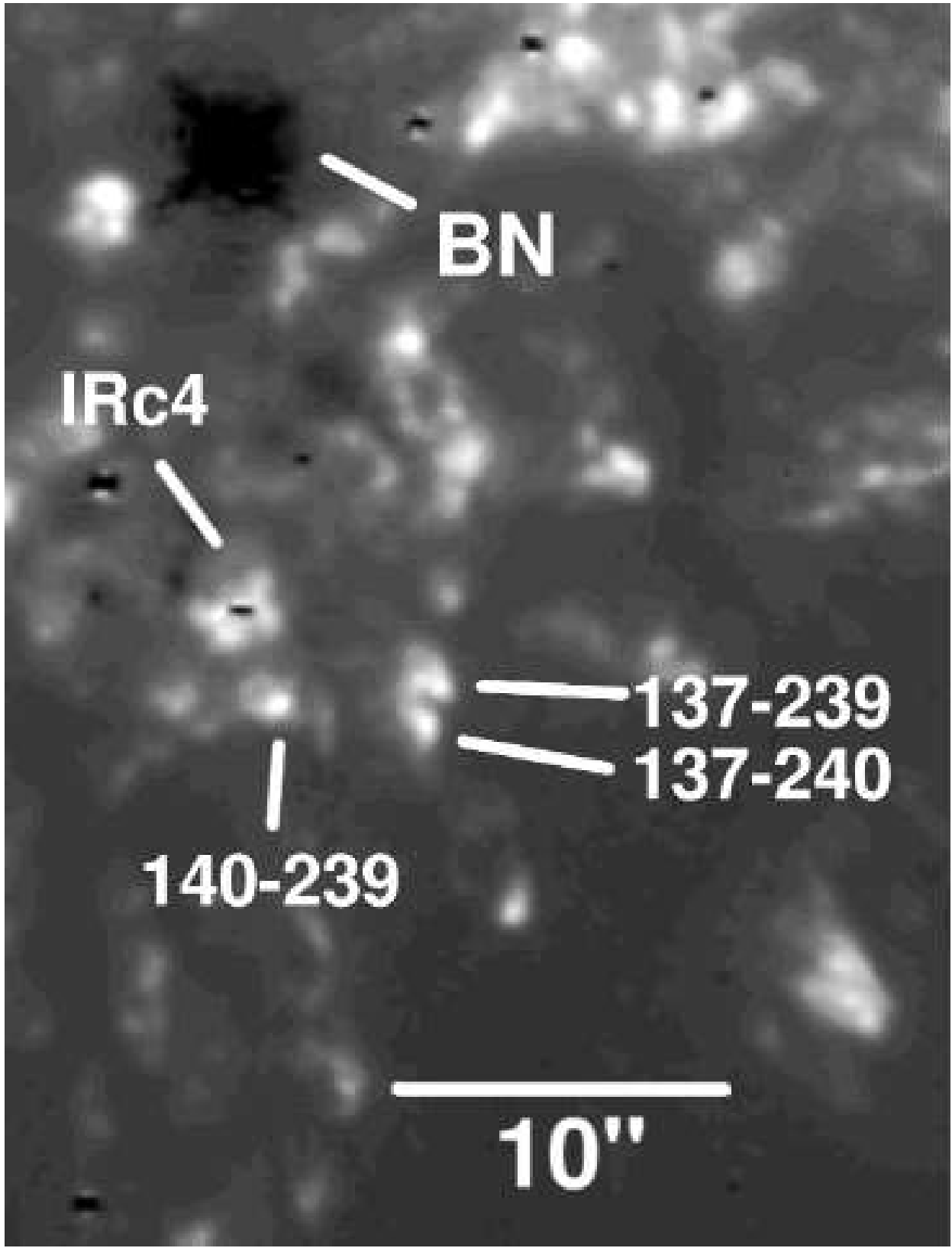}
\caption{H$_2$ 1--0 S(1) image of 137--239, 137--240, and 140--239. 
The maximum brightness in these features is 1.1$\times10^{-2}$ ergs s$^{-1}$cm$^{-2}$sr$^{-1}$. 
North is up and east is to the left.
\label{fig7}}
\end{figure}
\clearpage
\subsubsection{142--240, 143--239, and 144--237}

These three objects are shown in H$_2$ 1--0 S(1) emission in Figure 8. 
142--240 is a blunt, bow-shaped object southeast of the star at the head of IRc4. 
It is bright in H$_2$ and \ion{Fe}{2}, and much fainter in \ion{S}{2} and 
\ion{O}{1} \citep{Setal07}.  
The \ion{Fe}{2} emission \citep{Setal07} is more extended than the H$_2$, 
suggesting these transitions sample different regions.  
For 142--240, we deduce a C shock velocity of 42 km\,s$^{-1}$ 
and a pre-shock density of $10^{4.3}$ cm$^{-3}$. 
142--240 is accompanied on its eastern side by a fainter, 
larger bow-shaped region of \ion{Fe}{2} emission - 143--239 - 
which is not seen in \ion{S}{2} and \ion{O}{1} \citep{Setal07}. 
From its H$_2$ emission, which is blue-shifted \citep{SB07}, 
143--239 appears to arise from a J-shock: 
from Figure 4, the velocity and density would be in excess of 100 km\,s$^{-1}$ 
and $10^{7}$ cm$^{-3}$ respectively. 
A density this high might be expected to produce water masers, 
and indeed \citet{GWVJS} find a water maser 
within the error box of 143--239. 
143--239 connects to 144--237 - 
a bright knot of \ion{Fe}{2} \citep{Setal07} and H$_2$ emission 
4.\arcsec 4 to the northeast of 142--240.
Similar to 143--239, 144--237 exhibits blue-shifted emission \citep{SB07}, with  
a deduced shock velocity of 38 km\,s$^{-1}$ and a pre-shock density of $10^{4.3}$ cm$^{-3}$. 
There is also faint \ion{Fe}{2} emission slightly ($\sim 0.''6$) north 
of the H$_2$ emission \citep{Setal07}. 
If this \ion{Fe}{2} is associated with 144--237, 
it may be that 144--237 is moving in the direction of BN 
or that the bow shock is asymmetric. 
\clearpage
\begin{figure}
\figurenum{8}
\plotone{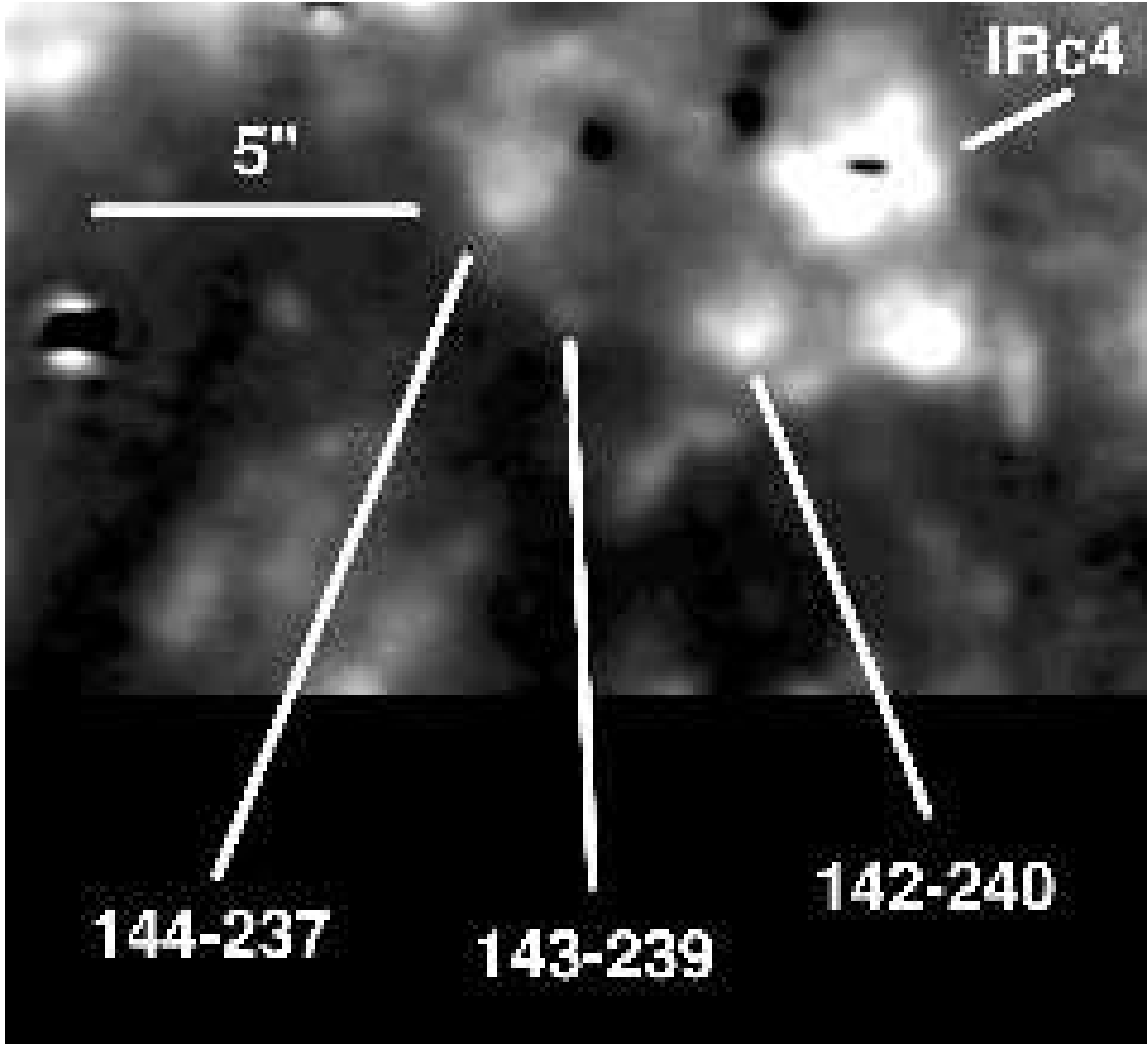}
\caption{H$_2$ 1--0 S(1) image of 142--240, 143--239, and 144--237. 
The maximum brightness in these features is 5.5$\times10^{-3}$ ergs s$^{-1}$cm$^{-2}$sr$^{-1}$. 
North is up and east is to the left.
\label{fig8}}
\end{figure}
\clearpage
\subsubsection{152--229 and 143-225}

\citet{Ch97} identified 152--229 as a bow shock - their bullet \#6. 
Our H$_2$, \ion{Fe}{2}, and 2.15 \micron\ continuum emission is shown in Figure 9. 
\citet{OHBM97} found that \ion{S}{2} emission 
in this region is blue-shifted. 
\citet{GKCFLPRL} found a peak velocity of -21 km\,s$^{-1}$ 
and a displacement of 0.\arcsec 2 between the emission peak 
and the location of the maximum velocity. 
From the direction of this displacement, 
they deduced that the shock is propagating towards BN 
(roughly between 2 and 3 o'clock in Figure 9). 
In disagreement with \citet{GKCFLPRL}, 
\citet{D02} found from the proper motion of the \ion{S}{2} emission, 
that 152--229 is moving slightly north of east - away from BN/IRc2 - 
with a transverse velocity of 50 km\,s$^{-1}$. 
\citet{Ch97} found 152--229 to have a FWZI of 110 km\,s$^{-1}$.  
The estimated shock velocity and pre-shock density are 
27 km\,s$^{-1}$ and $10^{5.0}$ cm$^{-3}$. 
\citet{Paper1} noted that the H$_2$ knot has a pointed cap of \ion{Fe}{2} emission 
on the southeast side of the object (the blue arc in Figure 9),
away from the putative exciting source (BN/IRc2) of the outflow.  
The positioning and morphology strongly suggests a bow shock in which 
strong shocks producing the \ion{Fe}{2} emission form on the leading surface of the bow 
while weaker shocks producing H$_2$ emission form behind it.  
The cap is also seen in the \ion{S}{2} and \ion{O}{1} images of \citet{OHBM97} 
and is possibly also visible in high-velocity \ion{S}{2} emission 
\citep{OHLWBRA97}; this may be the unlabelled \ion{S}{2} knot northeast of 147--234.  

\citet{Ch97} also identified 143--225 as a bow shock - their bullet \#8, 
with a a FWZI of 140 km\,s$^{-1}$. 
143--225 is among the most blue-shifted features in the outflow \citep{SB07} and 
may be a bow shock approaching us at a low angle. 
The H$_2$ emission is shown in Figure 10. 
The deduced shock velocity and pre-shock density are 
27 km\,s$^{-1}$ and $10^{5.8}$ cm$^{-3}$. 
The shape suggests that it is a bow shock pointed slightly north-northwest, 
which would mean the origin of the feature would be somewhere to the south-southeast - 
roughly opposite to the direction of BN. 

\clearpage
\begin{figure}
\figurenum{9}
\plotone{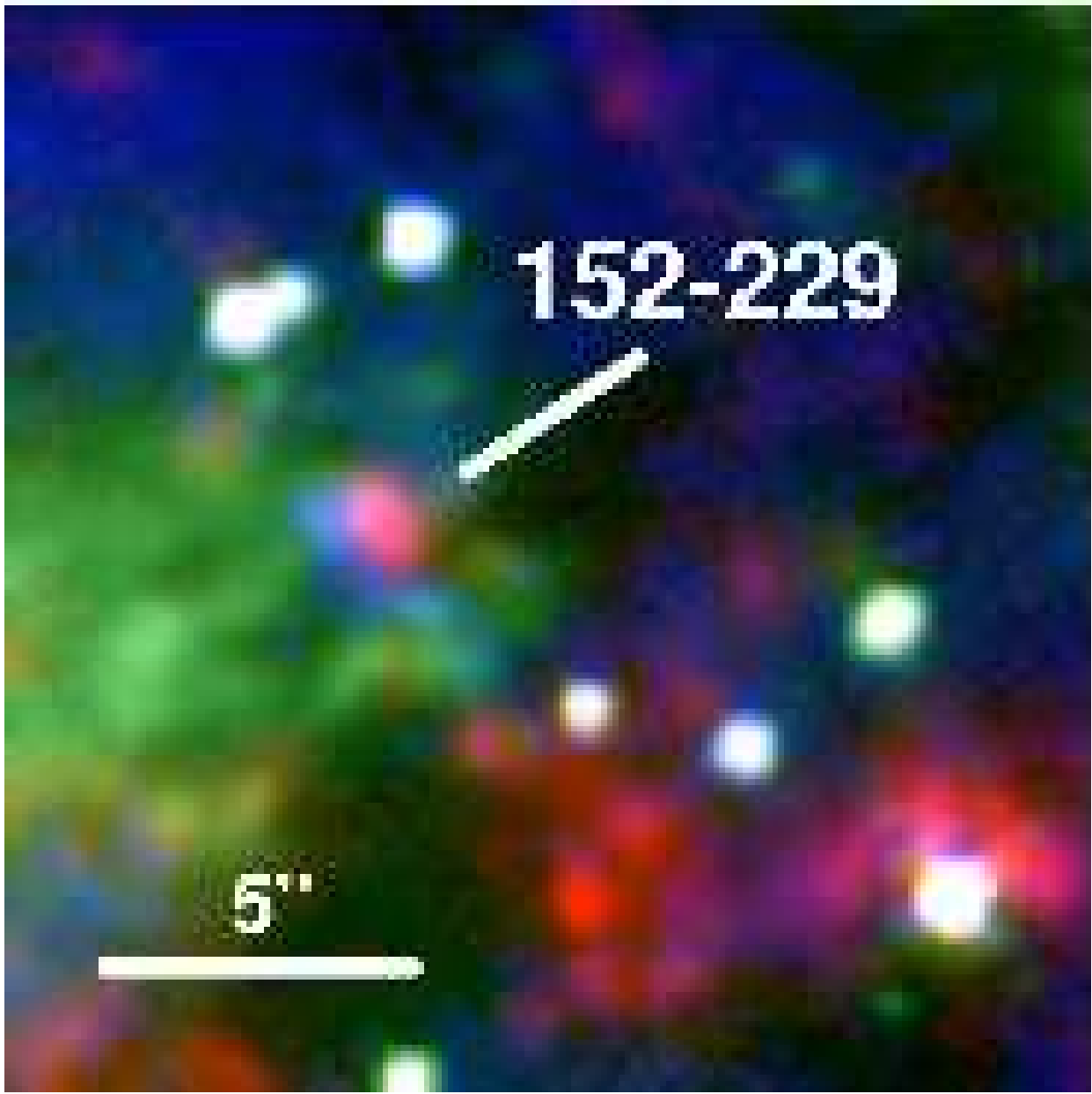}
\caption{False color image of 152--229.
Red is H$_2$ 1--0 S(1), green is the 2.15 \micron\ continum, and blue is \ion{Fe}{2}. 
North is up and east is to the left.
\label{fig9}}
\end{figure}
\clearpage

\begin{figure}
\figurenum{10}
\plotone{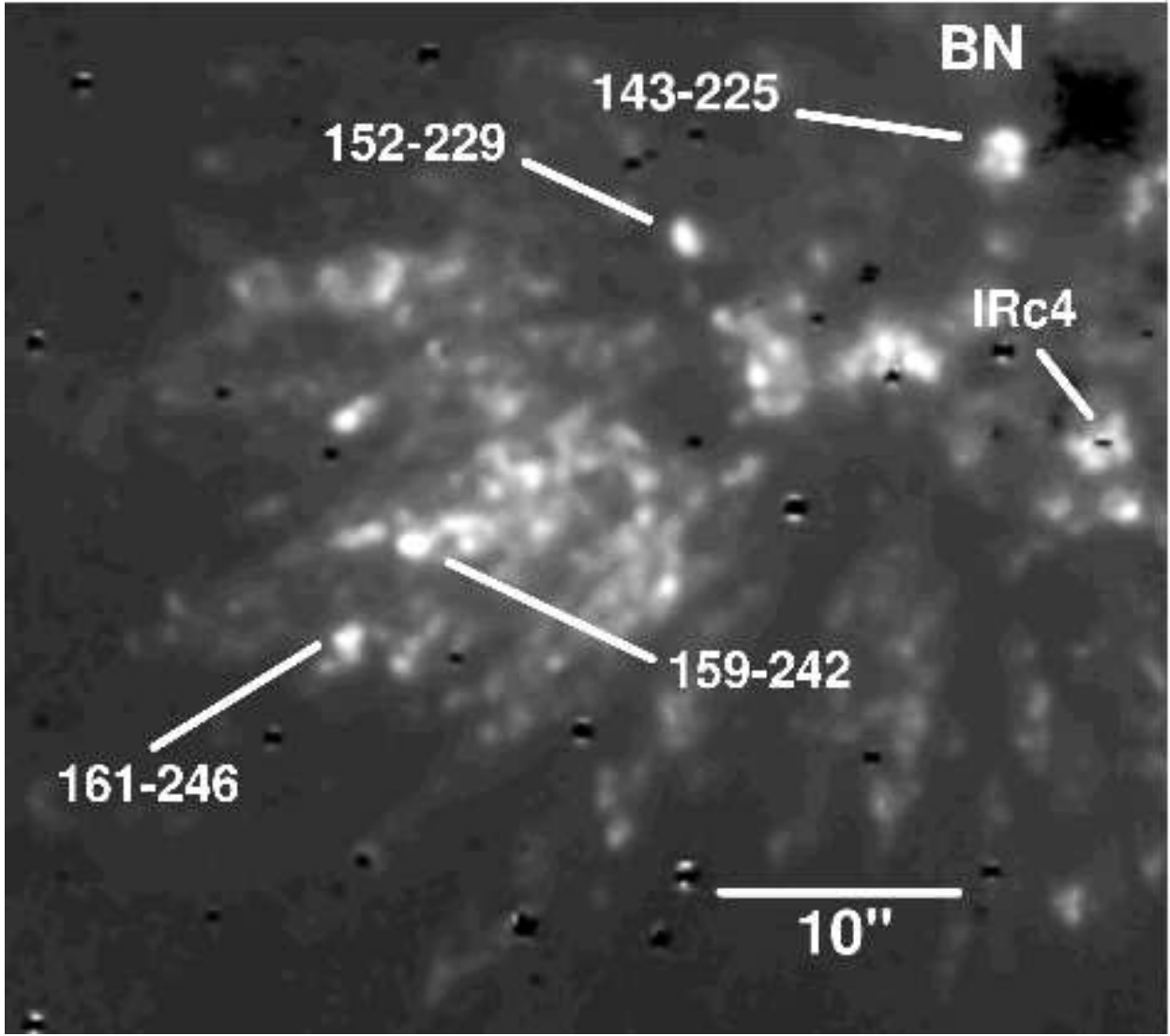}
\caption{H$_2$ 1--0 S(1) image of 143--225, 152--229, 159--242 and 161--246. 
The maximum brightness in these features is 1.8$\times10^{-2}$ ergs s$^{-1}$cm$^{-2}$sr$^{-1}$. 
North is up and east is to the left. 
\label{fig10}}
\end{figure}
\clearpage

\subsection{Knots}

\subsubsection{HH 208}

HH 208, approximately $7''$ west of BN, 
was first discovered by \citet{AT84}. 
The \ion{S}{2} and \ion{O}{1} HST images of \citet{OHLWBRA97} 
clearly show a number of small features. 
Based on \ion{S}{2} and \ion{O}{3} images, 
\citet{OHBM97} identified three knots in HH 208. 
The detection of optical features suggests that 
the extinction to HH 208 is lower than to other H$_2$ features, 
implying it lies more in the foreground. 
Figure 11 shows our H$_2$, \ion{Fe}{2}, and continuum images, 
along with our knot identifications. 
The HH 208 H$_2$ emission takes the form of discrete clumps, 
whereas the adjacent H$_2$ emission has a more fingerlike appearance.  
It is difficult to discern what process has created this collection of features.  

Knot A is the faintest of the 2.12 \micron\ H$_2$ knots, 
but is the original HH 208 - seen in both \ion{S}{2} and \ion{O}{1}. 
\citet{OHLWBRA97} showed images of knot A in several filters, and 
noted that a line drawn 
through HH 208 and HH 208NW terminates near the proplyd 154--240.  
They suggested that 154--240 may be the source of HH~208, but the line drawn
(which is symmetric through HH 208NW but not through the rest of the object)
also falls near IRc2 and radio sources ``I'' and ``n''.
\citet{D02} found no net proper motion of knot A. 
The structure of the bright core of knot A did change 
in a disorganized fashion between 1995 and 2000, 
which corresponds to motions over a range of about 50 km\,s$^{-1}$. 
Based on the high-velocity, blue-shifted emission lines, 
they further suggested that HH 208 is moving almost directly at us, 
rather than being connected to 154--240. 
In our data, knot A is the most extreme position - being beyond our grid of C-shock models. 
It is certainly higher velocity than any other HH 208 location, 
but may either be a high density ($>10^7$ cm$^{-3}$) J-shock, 
or a low density ($<10^4$ cm$^{-3}$) C-shock. 
The position of knot B, 
serving as a bridge between knot A and the rest of the object, 
is suggestive of a relationship between the forbidden-line and H$_2$ emission. 
At 42 km\,s$^{-1}$ and $10^{4.5}$ cm$^{-3}$, 
knot B has the second lowest density after knot U. 

\clearpage
\begin{figure}
\figurenum{11}
\plotone{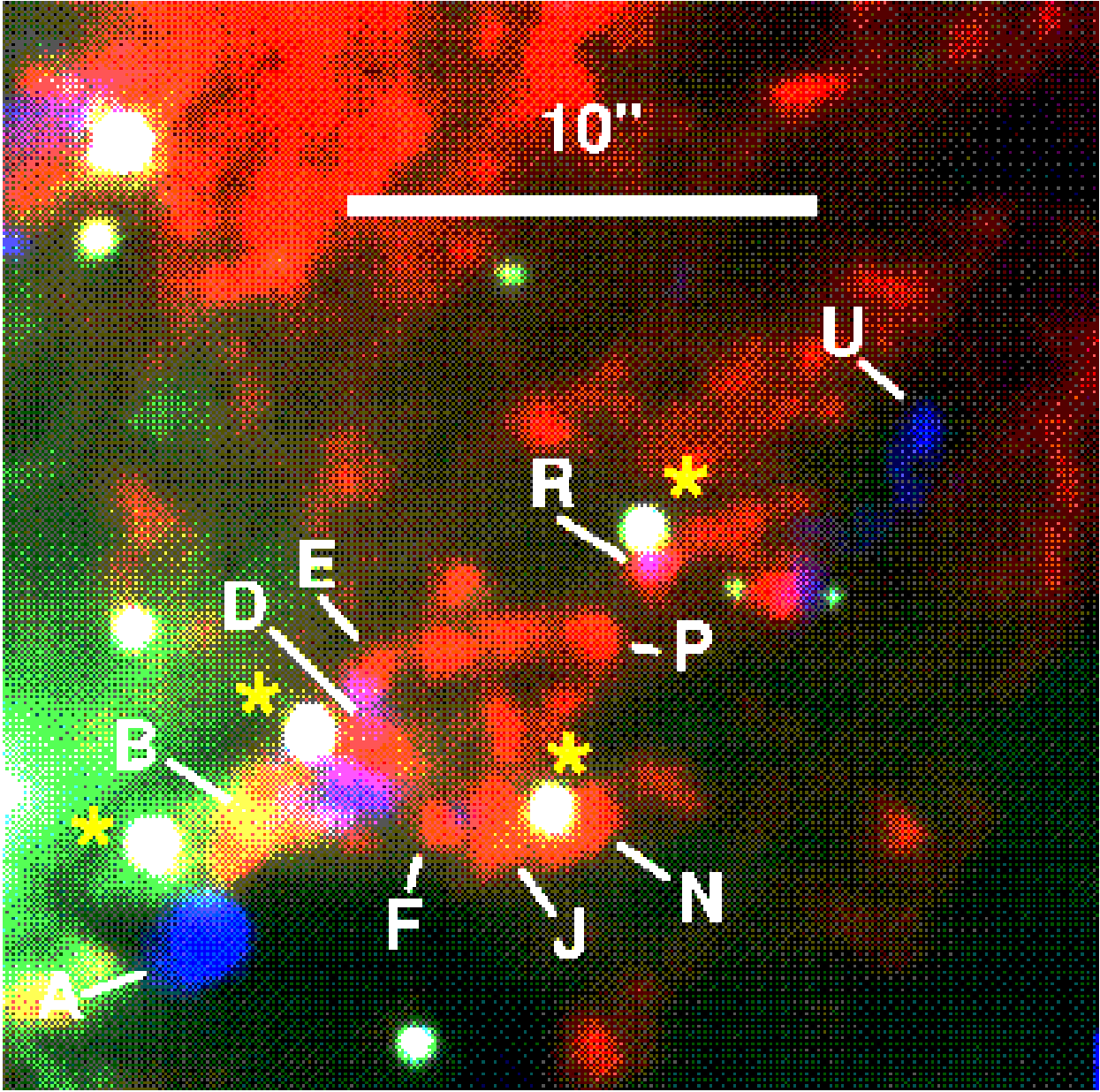}
\caption{False color image of HH 208 with features 
discussed in this paper labelled.
Red is H$_2$ 1--0 S(1), green is the 2.15 \micron\ continum, 
and blue is \ion{Fe}{2} 1.64 \micron. 
The gold star symbols denote stars. 
North is up and east is to the left.
\label{fig11}}
\end{figure}
\clearpage

The arrangement of knots D-E-P-N-J-F is suggestive of
the ``ring'' that \citet*{SMT97} pointed out 
in the northeastern edge of the Orion H$_2$ emission.  
The HH208 knots have deduced shock parameters in the 27--40 km\,s$^{-1}$ and 
$10^{4.5}$ -- $10^{5.2}$ cm$^{-3}$ ranges, with no clear pattern.
However, there may be some excitation gradient with distance from B.
Around knots B-D-E, the purple features show 
where the H$_2$ and the \ion{Fe}{2} emission coincide 
and the purple disappears beyond E. 
Since forbidden line emission arises from fast J-shocks, 
in these regions at least, 
the spatial coincidence of H$_2$ and \ion{Fe}{2} emission 
may be inconsistent with our general result that 
the H$_2$ emission in HH 208 arises in C-shocks. 
The knot of high-velocity \ion{S}{2} emission designated HH 208NNW by 
\citet{OHBM97} corresponds to the \ion{Fe}{2} emission 
we find accompanying H$_2$ knots D-E-F; 
this emission can also be seen in combined \ion{S}{2} and \ion{O}{1} emission in Figure 2 
of \citet{OHLWBRA97}. 
Farther away, knots J and N together form Object \#11 of \citet{Ch97}, 
one of the regions from which they detected discrete high velocity H$_2$ emission. 
Knot P shows neither forbidden line emission nor high-velocity H$_2$ emission.

Knots R and U together form HH 208NW \citep{OHBM97}. 
Knot R shows optical forbidden line emission, including  
high-velocity, blue-shifted \ion{S}{2} emission \citep{OHBM97}. 
The motion of R (129--216) in the plane of the sky is 49 km\,s$^{-1}$, 
and of U (126--214) is 65 km\,s$^{-1}$ \citep{D02}, both roughly to the west on a path 
that would have recently traversed the B-D-E-N-J-F ring. 
These proper motions put both these objects in the vicinity of BN/I/n 
around 1000 years ago \citep{D02}, 
consistent with many other H$_2$ features in the BN region, 
but significantly earlier than the 500 year old BN/I/n break-up \citep{R05, G05}. 
In our data, knot R shows the highest pre-shock density ($10^{5.5}$ cm$^{-3}$) 
but the lowest shock velocity (26 km\,s$^{-1}$) in HH 208 - 
suggesting the H$_2$ is not in the same region which produces 
the high velocity \ion{S}{2} emission. 
Knot U ties with B for the second highest velocity (42 km\,s$^{-1}$) and 
has the lowest density ($10^{4.0}$ cm$^{-3}$). 
Although they may be unrelated to the B-D-E-N-J-F ring, 
knot U (or perhaps knot R) may be faster-moving material - a wind or knot - 
that impacted the ambient material to create the ring. 
Knot A could be a faster moving section of the expanding ring which is moving towards us. 
The ambient material would then have been a local H$_2$ clump 
or even the core of a single, low mass star forming region.

\subsubsection{159--242 and 161--246}

159--242 and 161--246 - in the southwestern lobe of the outflow - 
are part of OMC Pk2 \citep{B78}. 
\citet{GKCFLPRL} found 159--242 (their object 6) to have 
a peak velocity of +11 km\,s$^{-1}$ and 161--246 (the western half of this knot
is their object 19) to have a peak velocity of -15 km\,s$^{-1}$. 
Applying shock models to their 2.12 \micron\ 1--0 S(1) flux measurements, 
\citet{V01} derived a pre-shock density of $\sim 10^6$ cm$^{-3}$, 
yielding a mass in 161--246 of 0.1 to 0.15 M$_\odot$, 
making it the most massive clump in their field.  
This led them to suggest that the 
clump is a candidate site for low-mass star formation.  
\citet{kgfclvp} expanded upon that work by including 2-1 S(1) images; 
revising their shock models in light of the new data led them to conclude 
that the density is an order of magnitude {\it greater} than \citet{V01} calculated.  
\citet{kgfclvp} also concluded that the 2-1 S(1)/1-0 S(1) flux ratio 
included a contribution due to radiative excitation from $\theta^1$C Ori, 
as well as to shocks.  
They were unable to reproduce both brightness and line ratios 
with a single type of shock, 
and therefore suggested that the shock contribution to the emission is composed of 
C-shocks in the interior of the clump, with J-shocks on the exterior. 
Our H$_2$ emission is shown in Figure 10. 
From our analysis, 159--242 and 161--246 are well fit by shock velocities of 
41 and 42 km\,s$^{-1}$ and much lower pre-shock densities 
of $10^{4.5}$ and $10^{4.4}$ cm$^{-3}$. 
This suggests that the material producing the 2.12 \micron\ 1--0 S(1) emission 
is insufficient to support even low mass star formation. 
We do find that the line ratio is about 50\% higher 
in a small region 0.9\arcsec\ east of the maximum H$_2$ emission in 159--242, 
implying a higher shock velocity and lower density there, 
consistent with the exterior J-shock proposed by \citet{kgfclvp}. 

\section{Summary}

From 0.\arcsec 2 (90 AU) angular resolution HST NICMOS narrowband images of OMC-1, 
which resolve individual shocks, 
we estimate the brightnesses of H$_2$ transitions at 1.89 and 2.12 \micron\ for 23 features. 
A comparison of the data with shock models shows 
that most of the data cannot be fitted by J-shocks, 
but are well matched by C-shocks 
with shock velocities in the range of 20--45\,km\,s$^{-1}$ 
and preshock densities of $10^{4} - 10^{6}$ cm$^{-3}$. 
The narrow range of shock velocities is not surprising since both  
slower C-shocks and faster J-shocks produce weaker H$_2$ emission 
and would not have been detected. 
Although there are many shock features in the OMC-1 region, 
most of the features appear to be well-characterized 
by a limited range of shock velocities and preshock densities, 
supporting the possibility of a common origin. 
Additionally, these values confirm the findings of larger beam studies, 
which averaged over a number of individual shocks. 
Two objects  - 143--239 and HH208A - are possibly due to J-shocks 
and the former does coincide with a known water maser. 
Optical forbidden line measurements of some features in HH 208 
require fast J-shocks for excitation; 
we cannot explain this apparent discrepancy. 

\acknowledgements

We wish to thank Janet Simpson, Robert Rubin, and an anonymous referee for careful readings of 
and helpful comments on the manuscript. 
A.S.B.S. acknowledges support from NASA/Ames Research Center Research Interchange 
grants NCC2-647 and NCC2-1134 to the SETI Institute.

\end{document}